\documentstyle[preprint,eclepsf]{jpsj}

\def\runtitle{Antiferromagnetic Phases of One-Dimensional
Quarter-Filled Organic Conductors}
\def\runauthor{Hitoshi {\sc Seo} and Hidetoshi {\sc Fukuyama}}

\def\lsim{\lower -0.3ex \hbox{$<$} \kern -0.75em \lower 0.7ex \hbox{$\sim$}}
\def\gsim{\lower -0.3ex \hbox{$>$} \kern -0.75em \lower 0.7ex \hbox{$\sim$}}
\def\S{\rm S}

\title{Antiferromagnetic Phases of\\
 One-Dimensional
Quarter-Filled Organic Conductors}
\author{Hitoshi {\sc Seo}\footnote{E-mail: hseo@watson.phys.s.u-tokyo.ac.jp} 
and Hidetoshi {\sc Fukuyama}}
\inst{Department of Physics, Faculty of Science, University of Tokyo, Tokyo 113}

\recdate{\today}

\abst{The magnetic structure of 
antiferromagnetically ordered phases of quasi-one-dimensional
organic conductors is studied theoretically at absolute zero
based on the mean field approximation to the quarter-filled band 
with on-site and nearest-neighbor Coulomb interaction.
The differences in magnetic properties between the antiferromagnetic phase of 
(TMTTF)$_2$X
and the spin density wave phase in (TMTSF)$_2$X are seen to be due to 
a varying degrees of roles played by the on-site Coulomb interaction.
The nearest-neighbor Coulomb interaction introduces charge
disproportionation, which has the same spatial periodicity as the 
Wigner crystal, accompanied by a modified antiferromagnetic phase.
This is in accordance with the results of experiments by Nakamura {\em et al} 
on (TMTTF)$_2$Br and (TMTTF)$_2$SCN.
Moreover, the  
antiferromagnetic phase of (DI-DCNQI)$_2$Ag is predicted
to have a similar antiferromagnetic spin structure.}
\kword{TMTCF, nearest-neighbor Coulomb interaction, mean field approximation,
antiferromagnetic spin ordering, charge disproportionation, DCNQI}

\begin{document}
\sloppy
\maketitle

Quasi-one-dimensional organic conductors which have
a quarter-filled (or 3/4-filled) band 
such as (TMTCF)$_2$X, (DCNQI)$_2$M(M=Ag,Li), MEM-(TCNQ)$_2$ exihibit a variety of 
phases. In this work, the microscopic origin of this variety
was theoretically searched for with special emphasis on 
(TMTCF)$_2$X and (DI-DCNQI)$_2$Ag. 

\begin{figure}
\begin{center}
\epsfile{file=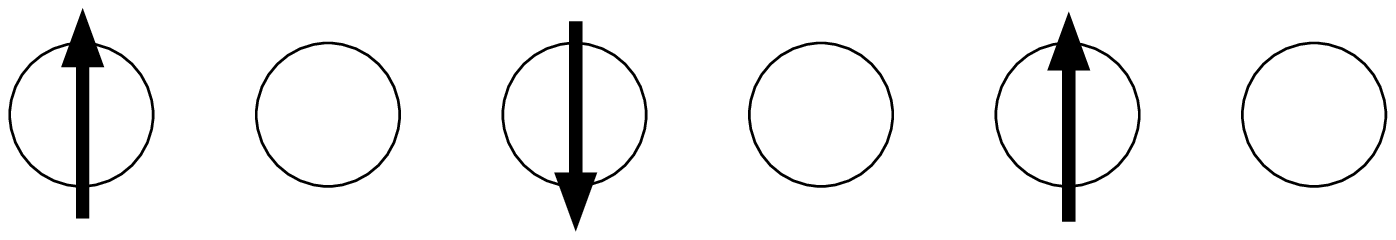,height=1cm}
\end{center}
\caption{Spin structure of (TMTTF)$_2$Br and (TMTTF)$_2$SCN in the 
 TMTTF chain, indicated by NMR experiments.\cite{Nakamura,Nakamura2}
 Circles represent
 TMTTF molecules, and arrows pointing upward (downward) represent places
 where the up (down) spin appears. }
\label{spinstruc}
\end{figure}

(TMTCF)$_2$X, where X=PF$_6$, ClO$_4$, Br, NO$_3$, etc. is a generic name
for the compounds (TMTSF)$_2$X, also called Bechgaard salts, 
and their sulphur analogs (TMTTF)$_2$X.\cite{review} 
(TMTSF)$_2$X salts have metallic room temperature resistivity and 
usually undergo a spin density wave (SDW) transition at
10-12K which is caused by the nesting of its Fermi surface.
The SDW in (TMTSF)$_2$PF$_6$, for example, 
has an incommensurate wave number 
(0.5, 0.20-0.24, 0-0.06) and an amplitude of 
0.08 $\mu_B$/molecule.\cite{Takahashi,Delrieu}
Application of 
hydrostatic pressure supresses the SDW transition and gives rise to
superconductivity.\cite{Jerome} 
In contrast, (TMTTF)$_2$X salts are semiconducting with a shallow minimum in 
resistivity at 250-100 K. 
This paramagnetic insulating phase has been understood 
as a Mott insulator due to electronic correlation,\cite{Barisic,Emery,Torrance}
because the one-dimensional (1-D) band of the TMTCF chain 
is effectively half-filled
because of the dimerization of molecules.
The ground states of the members of the 
(TMTTF)$_2$X family are either spin-Peierls
dimerized or antiferromagnetic states. 
In the case of (TMTTF)$_2$Br and (TMTTF)$_2$SCN, 
the ground state is an AF phase with a commensurate
wave number and magnetic moment at ambient pressure 
of (1/2, 1/4, 0) and 0.14 $\mu_B$/molecule, respectively,
as disclosed by Nakamura {\em et al.}\cite{Nakamura,Nakamura2} 
The line shape observed in these two experiments suggests that the 
microscopic spin structure in these salts is as shown in
Fig. \ref{spinstruc}.\cite{Takahashi_pc}
Furthermore, in (TMTTF)$_2$Br under pressure 
the existence of an AF state with 
an incommensurate wave number has been indicated\cite{Klemme}. 
This incommensurate phase can be considered to be similar to the SDW 
phase in (TMTSF)$_2$X.

In order to study the effect of the Coulomb interaction
in TMTCF salts, 
we consider a 1-D chain of TMTCF molecules.
It is expected that a reasonable description
will be provided by the 1-D dimerized extended Hubbard model
defined by the Hamiltonian

\begin{eqnarray}
H=&t_1& \sum_{i\ even,\sigma} \left( a^{\dagger}_{i\sigma}%
    a_{i+1\sigma}+h.c.\right)
    +t_2 \sum_{i\ odd,\sigma} \left( a^{\dagger}_{i\sigma}a_{i+1\sigma}
    +h.c. \right) \nonumber\\
    &+&U\sum_{i}n_{i\uparrow}n_{i\downarrow}+V\sum_i n_in_{i+1}, 
\label{eqn:Hamil}
\end{eqnarray}
where $\sigma$ is a spin index and takes $\uparrow$ and $\downarrow$,
$n_{i\sigma}$ and  $a^{\dagger}_{i\sigma}$ ($a_{i\sigma}$) denote 
the number operator and the creation (annihilation) operator for the 
electron of spin $\sigma$ at the $i$th site
and $n_i=n_{i\uparrow}+n_{i\downarrow}$.
$U$ and $V$ are the on-site and nearest-neighbor
Coulomb interactions, respectively. 

We treat $U$ and $V$ in the mean field (MF) approximation
and follow the treatment
by Kino and Fukuyama,\cite{Kino,Kino2} 
who have elucidated that the variety of ground states of several BEDT-TTF compounds
can be understood in terms of the on-site Coulomb interaction and the dimerization 
as key factors. 

We assume that 
four molecules make the basic lattice periodicity in the chain direction,
as is indicated by experiments.
Thus there are four different sites in a unit cell and the 
MF Hamiltonian in $k$-space is given by
\begin{full} 
\begin{eqnarray}
H^{MF}=\sum_{k\sigma}\left(\begin{array}{cccc}
a_{1k\sigma}\\
a_{2k\sigma}\\
a_{3k\sigma}\\
a_{4k\sigma} \end{array}\right)^{\dagger}
\left[h_{0}+U\left(\begin{array}{cc}
\begin{array}{cc}n_{1\bar{\sigma}} & \\
 & n_{2\bar{\sigma}}
\end{array} & 0 \\
0 & \begin{array}{cc}n_{3\bar{\sigma}} & \\
 & n_{4\bar{\sigma}}
\end{array}
\end{array}\right)\hspace{2.5cm}\right. \nonumber\\
\left.\hspace{2.5cm}+V\left(\begin{array}{cc}
\begin{array}{cc}n_2+n_4 & \\
 & n_1+n_3
\end{array} & 0 \\
0 & \begin{array}{cc}n_2+n_4 & \\
 & n_1+n_3
\end{array}
\end{array}\right)\right]
\left(\begin{array}{cccc}
a_{1k\sigma}\\
a_{2k\sigma}\\
a_{3k\sigma}\\
a_{4k\sigma} \end{array}\right),
\label{eqn:MFHamil}
\end{eqnarray}
\end{full}
where $\bar{\sigma}$ is opposite to $\sigma$ and $a_{\nu k\sigma}$
is the Fourier transform of $a_{i\sigma}$, i.e.
$a_{\nu{k}\sigma}=\frac 1{\sqrt{N_{\rm cell}}}\sum_{\alpha}
e^{iR_{\alpha\nu}k}a_{(\alpha\nu)\sigma}$ 
and N$_{\rm cell}$ is the total number of unit cells.
$\alpha$ and $\nu$ denote the cell number and 
the number of the molecule, respectively. The $h_0$ term is a matrix 
which comes from the first term of eq. (\ref{eqn:Hamil}).

Our calculation is carried out at $T=0$ and the electron density
is fixed at quarter filling, namely, two electrons per 4 molecules, 
since the band of (TMTCF)$_2$X is quarter-filled in terms of holes. 
It is supposed that the 
total spin moment in the unit cell is zero, i.e., the solutions we consider
are only paramagnetic or antiferromagnetic ones within a chain,
and possible ferro/ferrimagnetic states are excluded, which is reasonable
based on the results experiments. 
Actually in the entire range of parameters we show in the following,
no ferromagnetic states are stabilized.

\begin{figure}
\begin{center}
\epsfile{file=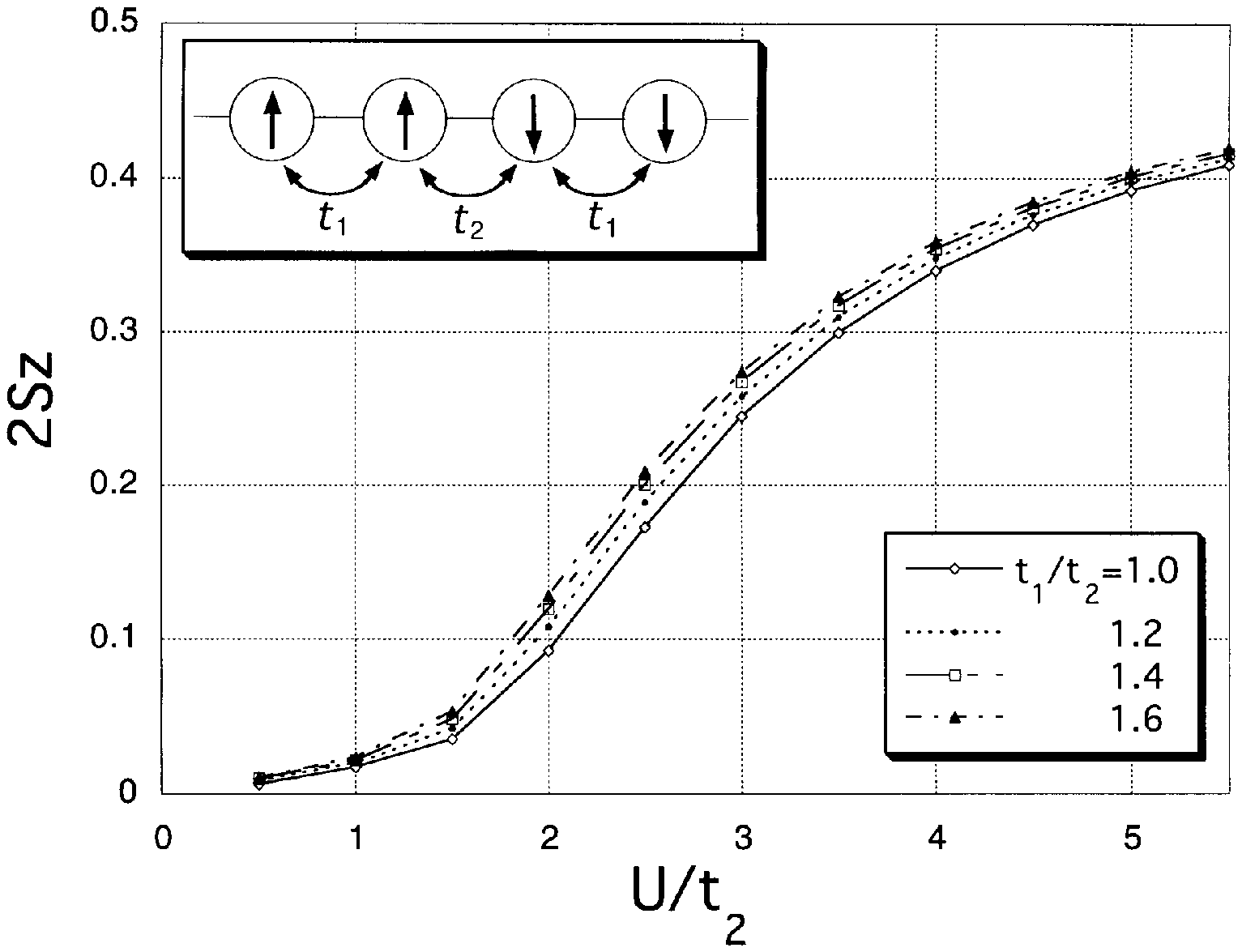,height=5.8cm}
\end{center}
\caption{U-dependence of absolute magnitude of spin moment
 per molecule, S$_Z$. }
\label{Sz-U}
\end{figure}

In the case of $V=0$, the results indicate that 
an AF state, whose spin configuration is as shown in
the inset of Fig. \ref{Sz-U}, is stabilized.
We set $t_1\geq t_2$ here,
so that the pairs of molecules (1--2) and (3--4) can be
considered as dimers.
The directions of the spin moments inside the dimers are the same 
and the AF ordering occurs between dimers, 
i.e. $\S_Z(1)=\S_Z(2)=-\S_Z(3)=-\S_Z(4)$ 
($\uparrow\ \uparrow\ \downarrow\ \downarrow$).
We have plotted the absolute value of
spin per molecule, S$_Z$,
as a function of $U/t_2$ 
for various values of $t_1/t_2$ in Fig. \ref{Sz-U}. 
It is seen that the magnitude of the spin moment increases rapidly when the on-site
Coulomb energy, $U$, increases and that the dependences
on $t_1/t_2$ are not so strong at least in the parameter range
investigated.
The wave number of this AF state is $2k_F$ which is identical to the 1-D nesting
vector, and this AF ordering affects the band structure so as to make the system an
insulator.
In the small $U$ region where the spin moment is small, 
the nested Fermi surface SDW picture
is valid, while in the large $U$ region
Mott AF (between the dimers) picture with larger spin moment 
would be more suited, although one cannot distinguish these
two pictures definitely within the MF approximation. 
 The electron density at each site remains 0.5 when $U$ is varied. 

\begin{figure}
\begin{center}
\epsfile{file=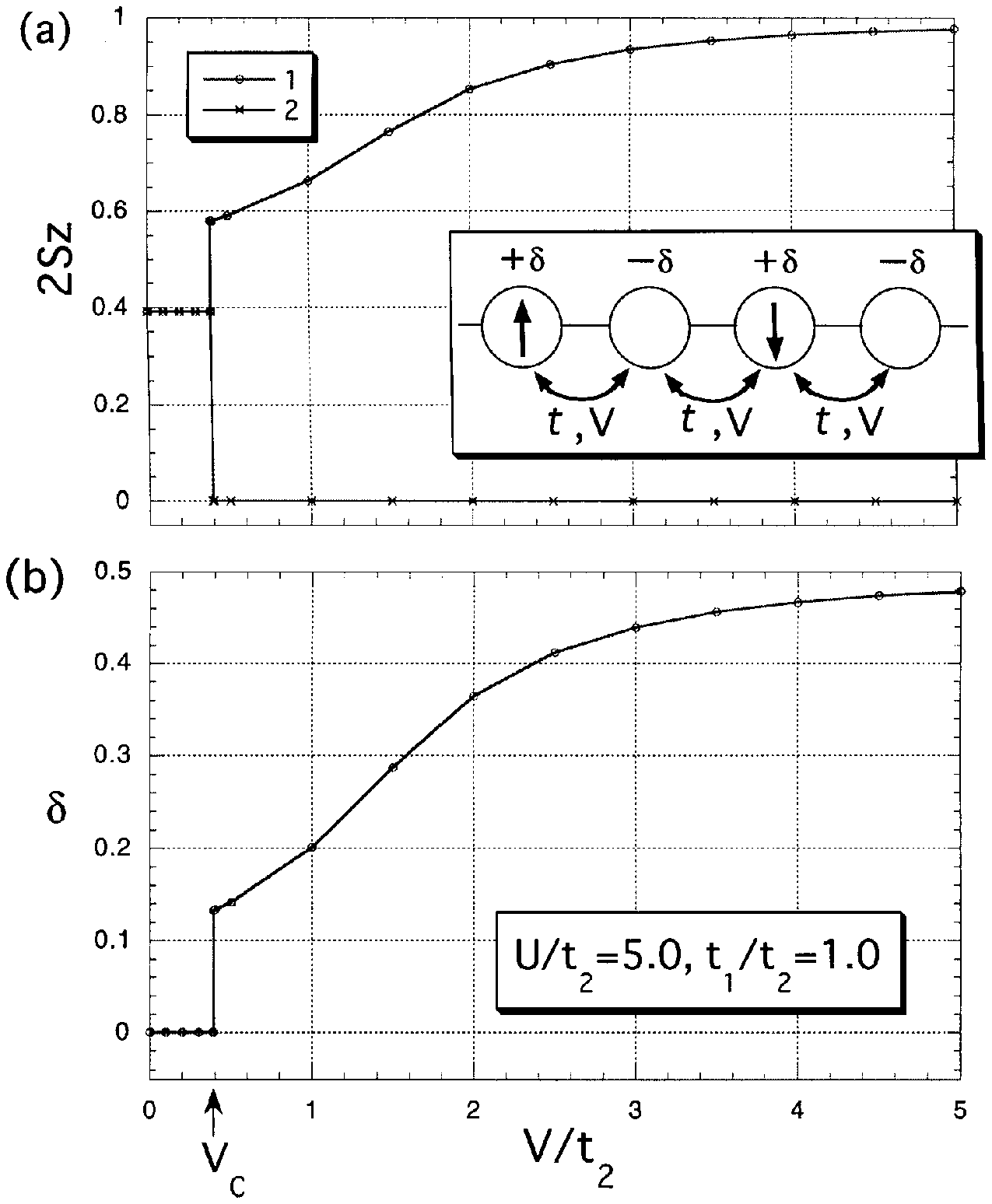,height=9.5cm}
\end{center}
\caption{V-dependence of absolute magnitude of spin moment per molecule, 
         $\S_Z$ (a)
         and charge disproportionation, $\delta$ (b) for $U/t_2=5.0,\ t_1/t_2=1.0$. } 
\label{Sz-V2}
\end{figure}

\begin{figure}
\begin{center}
\epsfile{file=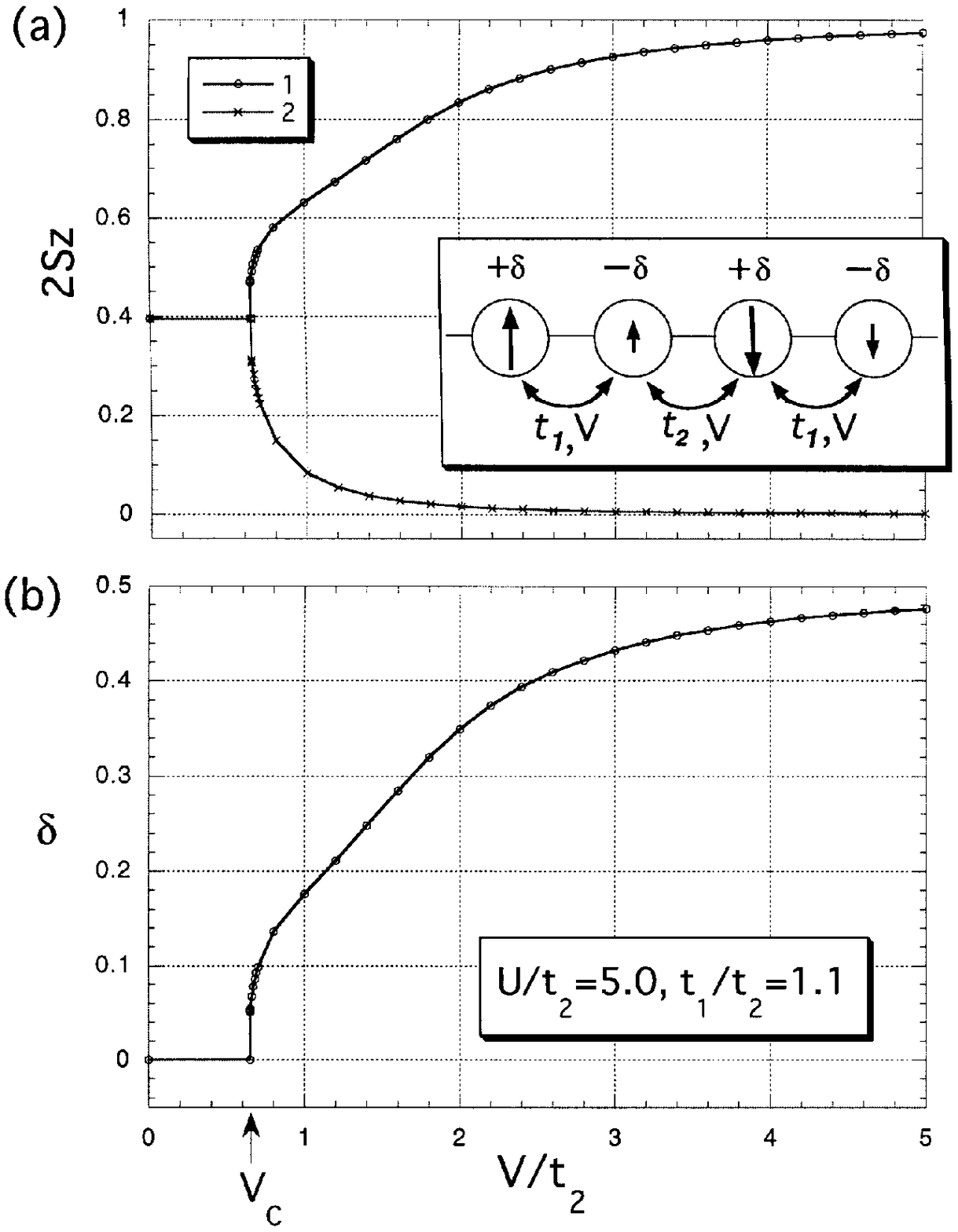,height=9.5cm}
\end{center}
\caption{V-dependence of absolute magnitude of spin moment per molecule, $\S_Z$ (a)
         and charge transfer, $\delta$ (b) for $U/t_2=5.0,\ t_1/t_2=1.1$. } 
\label{Sz-V4}
\end{figure}

In the presence of finite $V$,
a phase transition occurs between
two different types of AF states with a fixed value of $U$. 
We show in Figs. \ref{Sz-V2} and \ref{Sz-V4} the spin, S$_Z$, 
of molecules 1 and 2 for 
$t_1/t_2=1.0$ (without dimerization) and 1.1 (with dimerization)
with $U/t_2$ fixed at 5.0. 
It is seen that the charge disproportionation, $\delta$, is accompanied by
this transition.
We note that $\S_Z(1)=-\S_Z(3)$, $\S_Z(2)=-\S_Z(4)$ and
$n_1=n_3=0.5+\delta,\ n_2=n_4=0.5-\delta$, where $n_i$ is the electron density
at site `$i$'.

For comparison, we first analyze the results for $t_1=t_2$ (Fig. \ref{Sz-V2}),
which is the case without dimerization.
It is seen that there exists 1st-order 
transition at $V=V_c\ (V_c/t_2\simeq 0.392$
for $U/t_2=5.0)$ 
from the ($\uparrow\ \uparrow\ \downarrow\ \downarrow$) AF state to
another AF state, where the spin moment in molecules 2 and
4 is zero and only the up-spin in molecule 1 and the down-spin in the molecule 3
survive ($\uparrow 0 \downarrow 0$), as shown in an inset of Fig. \ref{Sz-V2}.
As regards the $V$-dependence of the spin moment, for $V\leq V_c$ the 
absolute value of the moment is not different from that in the case of $V=0$,
while for $V\geq V_c$, 
$\S_Z(1)=|\S_Z(3)|$ increases with increasing $V$. 
This ($\uparrow 0 \downarrow 0$) AF state is accompanied by the charge disproportionation,
i.e. the electrons on molecules 2 and 4 move to molecules 1 and 3.
This modulation of the charge is the $4k_F$ charge-density-wave (CDW)
whose amplitude is twice the value of $\delta$. 
This state can be considered as a Wigner crystal.

In the case of $t_1\neq t_2$ (Fig. \ref{Sz-V4}),
i.e. with dimerization,
the latter ($\uparrow 0 \downarrow 0$) AF state is modified.
There appear spin moments also on molecules 2 and 4,
and the absolute values of the magnetic 
moments on the two molecules within the dimer are different, as 
$\S_Z(1)\geq \S_Z(2)$ and $|\S_Z(3)|\geq |\S_Z(4)|$ 
($\uparrow{\scriptstyle \uparrow}\downarrow{\scriptstyle \downarrow}$), 
which is schematically shown in the inset of
Fig. \ref{Sz-V4}. 
We note that there exists a charge disproportionation 
with a magnitude similar to that in the case of 
$t_1=t_2$.
The value of $V=V_c$, where this phase transition from the
($\uparrow\ \uparrow\ \downarrow\ \downarrow$) state to the 
($\uparrow{\scriptstyle \uparrow}\downarrow{\scriptstyle \downarrow}$) 
state occurs, 
gets large when one increase the degree of dimerization 
$t_1/t_2$, and it appears that the 1st-order phase transition crosses over to 
a 2nd-order-like behavior as $t_1/t_2$ is increased.
Above $V_c$, the $V$-dependence of the spin moment of molecules 1 and 3
is similar to that in the case of $t_1=t_2$,
though that of molecules 2 and 4 falls 
rapidly toward zero as $V$ increases. 

In both cases of $t_1=t_2$ and $t_1\neq t_2$, the magnetic moment 
has a tendency to increase with increasing $U$ when $V$ and $t_1/t_2$ are fixed.

We now compare the results of the calculation and 
the experimental results, choosing the 
parameters of our theoretical model to fit the actual compounds.
In (TMTCF)$_2$X, 
the Coulomb interaction, $U$, on a TMTCF
molecule is considered to be almost the same value in both series and of order
1 eV.
Extended H\"uckel band calculations\cite{Mori,Grant,Mori2,Ducasse} show that
$t_2\simeq 0.2$ eV for TMTTF compounds and $t_2\simeq 0.3$ eV for 
TMTSF compounds.
Thus the ratio $U/t_2\simeq 5$ for TMTTF is larger than
the value $U/t_2\simeq 3$ for TMTSF, and the experimental results for 
the large spin moment observed in (TMTTF)$_2$Br and
(TMTTF)$_2$SCN compared to the SDW amplitude are consistent with
the present theory.

The value of $t_1/t_2$ is expected to be about 1.1 
in both (TMTTF)$_2$Br and
(TMTTF)$_2$SCN,\cite{Mori,Grant,Mori2,Ducasse} 
hence Fig. \ref{Sz-V4} applies.
The spin structure ($\uparrow 0 \downarrow 0$) suggested by the 
results of H-NMR 
experiments\cite{Nakamura,Nakamura2,Takahashi_pc} 
in (TMTTF)$_2$Br and (TMTTF)$_2$SCN, which is shown in 
Fig. \ref{spinstruc}, is apparently different from that given 
in Fig. \ref{Sz-V4}, but this apparent discrepancy
can be understood as follows. 
When the ($\uparrow{\scriptstyle \uparrow}\downarrow{\scriptstyle 
\downarrow}$) 
AF state is realized and 
the spin moment in molecules 2 and 4 is negligible compared 
to that in molecules 1 and 3, 
i.e. when $\S_Z(2)=|\S_Z(4)|\ll\S_Z(1)=|\S_Z(3)|$, 
the observed line shape will be indistinguishable
from that in the case of ($\uparrow 0 \downarrow 0$).
It is seen in  Fig. \ref{Sz-V4} that S$_Z(2)\ \lsim\  0.1\times $S$_Z(1)$
when $V/U\ \gsim \ 0.25$.
Such values of $V/U$ are consistent with the estimate by 
Mila,\cite{Mila} and the quantum chemistry calculations.\cite{Fritsch,Castet} 
Consequently we conclude that 
the long-range Coulomb interaction
plays an essential role in (TMTTF)$_2$Br and (TMTTF)$_2$SCN, and
probably in the other TMTTF series as well. 

The ($\uparrow{\scriptstyle \uparrow}\downarrow{\scriptstyle \downarrow}$) 
state is also possible in the presence of two-fold periodic potential,
instead of the nearest-neighbor Coulomb interaction. 
This case with two-fold periodic potential has been investigated in
another context.\cite{Tanemura}
It turned out, however, that to realize a state without $V$ 
where the spin moment in molecules 2 and 4 is 
negligible compared to that in molecules 1 and 3,
we need to introduce an unreasonably large two-fold
periodic potential.\cite{Seo}
Furthermore, the existence of such two-fold periodic potential
is not expected from the crystallography,\cite{Takahashi_pc,Kagoshima_pc}
because the anions
are located at the vertices of a parallelipiped and two TMTCF molecules
are placed centrosymmetrically in the central cavities. 

Next we will analyze the effect of pressure on (TMTCF)$_2$X. 
It is considered that the
bandwidth increases and the degree of dimerization decreases with increasing 
pressure, while $U$ will be hardly affected, then the reduced on-site
Coulomb interaction $U/t_2$ and the ratio $t_1/t_2$ 
will both decrease.  
The present theoretical results indicate 
that the magnetic moment decreases when
$U/t_2$ and/or $t_1/t_2$ decreases. Thus it is expected that 
the materials whose ground state is Mott AF at ambient pressure
will be transformed into an SDW as the pressure is increased. 
In (TMTTF)$_2$Br the transition between a Mott AF phase to an SDW phase
was actually observed.\cite{Klemme}

Finally we discuss the properties of (DCNQI)$_2$Ag,
which is quite similar to (TMTTF)$_2$X. 
A recent resistivity measurement\cite{Hiraki} on (DI-DCNQI)$_2$Ag 
has disclosed an insulating behavior in the entire temperature
range below 300 K. Below 200 K, a $4k_F$ CDW 
which is of the charge modulation type has been observed
in an NMR experiment,\cite{Hiraki2}
and an AF transition occurs at 5.5 K.\cite{Hiraki} 
On the other hand, (DMe-DCNQI)$_2$Ag is metallic down to 100 K, 
below which the system crosses over to insulating phase
accompanied by a $4k_F$ CDW,
as suggested by x-ray diffusive scattering.\cite{Moret}

Since (DCNQI)$_2$Ag systems have no dimerization along the 
DCNQI stack direction, i.e. $t_1=t_2(=t_\parallel)$, 
in contrast to the TMTCF family, 
which is quarter-filled, the existence of a 
Mott insulator produced by on-site Coulomb
interaction, $U$, is not possible, while a Wigner crystal
due to nearest-neighbor Coulomb interaction, $V$, is a candidate. 
Actually, the intrachain transfer energy $t_\parallel$ is estimated 
to be about 0.2 eV by first principles calculation,\cite{Miyazaki}
and the on-site Coulomb interaction $U$ is of the order of 1 eV, 
so $U/t_\parallel\simeq 5$, 
which is similar to the value for TMTTF salts.
Moreover, $V$ is expected to be quite large as in the case of TMTTF compounds.
Then the electronic structure of the ground state is
expected to be ($\uparrow 0 \downarrow 0$) AF state
as shown in Fig. \ref{Sz-V2}.
This consideration is consistent with the observed 
$4k_F$ CDW.\cite{Hiraki2,Moret}

We note here that, in spite of the capability of the exploration of the general
trends in a unified way so far demonstrated, 
the present MF scheme does not take account of fluctuations, 
and overestimates the stability of magnetic orderings.
In fact, the magnetic moments obtained in our MF
calculations are larger than the observed value
in TMTCF compounds.
For understanding of 
such more quantitative features, effects of the interchain transfer 
should be taken into account, which is under study.

In summary, we have studied the effects of on-site and 
nearest-neighbor Coulomb interaction 
in a quarter-filled 1-D band with and without dimerization, and found that 
various types of AF spin structure are stabilized. 
In the absence of $V$, the spin structure of the type 
($\uparrow\ \uparrow\ \downarrow\ \downarrow$) is the ground state
independent of the degree of the dimerization, and the charge density
is uniform. The staggered magnetization is larger for larger $U$ in general. 
On the other hand, once $V$ is larger than some critical value charge
disproportionation is realized which can be considered as a tendency 
toward formation of the Wigner crystal. The spin structure in this case is   
($\uparrow 0 \downarrow 0$) or 
($\uparrow{\scriptstyle \uparrow}\downarrow{\scriptstyle \downarrow}$) 
depending on without or with dimerization. 
These theoretical results explain experimental findings in 
(TMTTF)$_2$X and predict the AF spin structure in (DI-DCNQI)$_2$Ag. 
This is the first reported study which has clarified the 
unique roles of $V$ and dimerization in stabilization of characteristic 
spin structures in 1-D quarter-filled band. 

\acknowledgements
We thank H. Kino, T. Takahashi, S. Kagoshima, K. Hiraki,
K. Kanoda, D. Yoshioka and H. Kohno 
for useful discussions and suggestions. 
This work was financially supported by a 
Grant-in-Aid for Scientific Research on Priority Area ``Anomalous Metallic
State near the Mott Transition'' (07237102) from the Ministry of Education, 
Science, Sports and Culture.

\end{document}